\documentclass{article}

\usepackage{paralist}
\usepackage{times}
\usepackage{helvet}
\usepackage{courier}
\usepackage{graphicx}
\usepackage{caption}
\usepackage{subcaption}
\usepackage{color}
\usepackage{amssymb}
\usepackage{amsmath}

\newcommand{\addby}{{\mathsf{addrBy}}}
\newcommand{\subcon}{{\mathsf{subConc}}}
\newcommand{\conrate}{{storeAll}}  
\newcommand{\basicone}{{basicOne}}
\newcommand{\lowerrate}{{rate25\!f\!ps}}
\newcommand{\confi}{{Con\!f'\!d}}

\newcommand{\specialcell}[2][c]{%
  \begin{tabular}[#1]{@{}c@{}}#2\end{tabular}}

\newcommand{\entails}{\vDash}

\newcommand{\lpnot}{\mbox{not}\;\,}

\newcommand{\hif}{\leftarrow}

\newcommand{\causes}{\mbox{ causes }}
\newcommand{\lawif}{\mbox{ if }}

\newcommand{\triggers}{\mbox{ triggers }}
\newcommand{\defaults}{{\mathsf{defaults}}}

\newcommand{\aspindent}{\hspace*{.5in}}

\newcommand{\impactsneg}{\ \mbox{impacts}_{\mbox{neg}}\ }
\newcommand{\impactspos}{\ \mbox{impacts}_{\mbox{pos}}\ }

\begin{document}


\begin{center}
{\LARGE Ontology-Based Reasoning about the\\ Trustworthiness of Cyber-Physical Systems}
\end{center}

\bigskip\bigskip

\begin{center}
Marcello Balduccini\\ Saint Joseph's University, Philadelphia, PA, USA\\ {\tt marcello.balduccini@sju.edu}
\end{center}

\begin{center}
Edward Griffor\\ NIST, Gaithersburg, MD, USA\\ {\tt edward.griffor@nist.gov}
\end{center}

\begin{center}
Michael Huth\\ Imperial College London, London, UK\\ {\tt mhuth@imperial.ac.uk}
\end{center}

\begin{center}
Claire Vishik\\ Intel Corporation, Austin, TX, USA\\ {\tt claire.vishik@intel.com}
\end{center}

\begin{center}
Martin Burns\\ NIST, Gaithersburg, MD, USA\\ {\tt martin.burns@nist.gov}
\end{center}

\begin{center}
David Wollman\\ NIST, Gaithersburg, MD, USA\\ {\tt david.wollman@nist.gov}
\end{center}


\bigskip

\begin{abstract}
It has been challenging for the technical and regulatory communities to formulate requirements for trustworthiness of the cyber-physical systems (CPS) due to the complexity of the issues associated with their design, deployment, and operations.  The US National Institute of Standards and Technology (NIST), through a  public working group, has released a CPS Framework that adopts a broad and integrated view of CPS and positions trustworthiness among other aspects of CPS.  This paper takes the model created by the CPS Framework and its further developments one step further, by applying ontological approaches and reasoning techniques in order to achieve greater understanding of CPS. The example analyzed in the paper demonstrates the enrichment of the original CPS model obtained through ontology and reasoning and its ability to deliver additional insights to the developers and operators of CPS. 
\end{abstract}

\section{Introduction} 
\label{section:introduction}
\newcommand\blfootnote[1]{%
  \begingroup
  \renewcommand\thefootnote{}\footnote{#1}%
  \addtocounter{footnote}{-1}%
  \endgroup
}
\blfootnote{\emph{Official contribution of the National Institute of Standards and Technology; not subject to copyright in the United States. Certain commercial equipment, instruments, or materials are identified in this paper in order to specify the experimental procedure adequately. Such identification is not intended to imply recommendation or endorsement by the National Institute of Standards and Technology, nor is it intended to imply that the materials or equipment identified are necessarily the best available for the purpose.}
}
The cyber-physical systems (CPS) brought additional complexity to the computing environment. In addition to other requirements, the technologists now have to contend with the behavior and influence of the physical subsystem, creating an even greater need for an integrated context and the ability to reason about the application of the requirements.

The use of ontologically inspired modeling in computer science is not new. In fact, as Smith and Welty \cite{SmithWelty2001} point out, this approach has been used extensively in information systems science. Examples include conceptual modeling in the database development area or domain modeling in software engineering. Although these uses are separate from applying ontologies to knowledge engineering, there is a direct connection.

The creation of an extensive ontology is frequently a lengthy process. However, in this case, the authors had the advantage to rely on an extensive model already in existence. NIST hosted a Public Working Group on Cyber Physical Systems (CPS) with the aim of capturing input from those involved in CPS in order to define a reference framework supporting common definitions and facilitating interoperability between such systems. A key outcome of that work is the CPS Framework (Release 1.0) \cite{CPSFramework-vol1}. The framework proposes a means of supporting three Facets of a CPS life cycle: conceptualization, realization, and assurance of CPS through analytical lenses, called Aspects. In the framework, the Aspect named Trustworthiness, describes a number of related Concerns that deal specifically with the avoidance of flaws in Privacy, Security, Safety, Resilience and Reliability. The framework is extensible and supported with executable models, e.g. a UML model of Concerns and Aspects, and all three Facets and the interdependencies across the CPS life cycle. 
 
The CPS framework 
helps articulate 
the motivation for important requirements to be considered in building, composing, and assuring CPS. However, the CPS Framework currently does not offer a comprehensive model for 
reasoning over CPS artifacts and their dependencies.

In this paper, we develop a Conceptual Ontology for the Trustworthiness aspect that can be extended to other Aspects of the CPS Framework. 
We illustrate this approach with a case study, where the Conceptual 
Ontology is used to model the CPS from 
scenarios associated with a camera placed onto an autonomous car in order to support multiple aspects of decision making.

The model contains sufficient complexity to demonstrate the capabilities of the approach and how it can be scaled to the full CPS Framework. The case study includes, e.g., considerations such as Transduction (in which a CPS produces a physical signal that interacts with the Environment) and Influence (in which a CPS produces or receives a physical signal that brings about a state change of another CPS).  The objective is to demonstrate that an ontology-based approach can aid engineers in identifying and resolving important issues for design, implementation, and validation of CPS.

\noindent {\bf Intended Audience:} This paper is meant for both academic researchers and engineering professionals. For the former, it can stimulate more research in an area that urgently needs firm foundations for modeling and reasoning about the trustworthiness of CPSs. For the latter, it conveys the main ideas behind our approach and demonstrates that it can, in principle, be used in standard engineering and production practice.

\section{Related Work}
\label{section:related}

Ontology-Based Data Access (OBDA) systems (see e.g. \cite{InsurancePaper,bks15}) such as\ Ontop, allow for  semantic queries about an ontology to be interpreted over concrete data~--~using engines such as NoSQL, Hadoop, MapReduce and so forth. This is achieved through \emph{mappings} that mediate between the semantic layer of ontologies and the concrete data. Use of these maps can virtualize the concrete data graphs to those portions that are needed for evaluating the queries, improving scalability and semantically guiding data analytics, see e.g.\ \cite{InsurancePaper}. 
Our work is consistent with the use of OBDA to link to, and support, data analytics.

The Object Management Group has an Insurance Working Group that builds data models for that sector, informed by ontologies. Since ontologies can be composed, we may integrate such Insurance Ontologies as another important concern in the operation of CPS, particularly those related to infrastructure. 

For the Cybersecurity concern, there is a rich literature on graph-based attack models. Closest to our work are perhaps the Attack-Countermeasure Trees (ACT) by Roy, Kim, and Trivedi \cite{DBLP:journals/scn/RoyKT12}. An ACT specifies how (or how likely) an attacker can logically realize a specific goal in a IT system, even when faced with specific mitigation or detection measures. Leaves on trees are basic attack, detection or mitigation actions and the model assumes that basic attack actions are statistically independent.
Our approach is much wider in scope: it applies to CPS, is applicable to all concerns of the CPS framework and their dependencies~--~not just cybersecurity, and it can formulate and invoke inference rules of interest rather than relying on a static inference structure determined by a graph.

Our approach can be extended to \emph{quantitative} reasoning by interpreting queries and inferences as developed in this paper over the reals, rankings or other domains that allow a quantitative comparison. One may then generate answers to queries that are \emph{optimal} with respect to some metrics. The combination of physical (non-linear) interaction and logical (discrete or Boolean) interaction  of CPS make this a mixed-integer, non-linear optimization problem (MINLP) extended with logical inference. 
MINLP approaches can support a limited form of logic, e.g.\ through disjunctive programming \cite{balas:1975}. But these methods seem to struggle with supporting richer logics and inferences such as ``what-if'' explorations. We therefore seek support for both MINLP 
methods and logic reasoners. This need has already been recognized in the optimization community, we refer to \cite{ruthOnline17} for an overview, a discussion, and first results in addressing this need for Process Systems Engineering. The tool ManyOpt \cite{DBLP:journals/corr/DIddioH17} already provides such abilities but can only express polynomials as non-linear behavior. The notion of $\delta$-satisfiability~\cite{DBLP:conf/lics/GaoAC12} relaxes inequalities by up to some $\delta > 0$ in order to satisfy all constraints. This renders decidability for a rich theory including transcendental functions, with tool support \cite{DBLP:conf/cade/GaoKC13}. It would be of great interest to leverage this to optimization plus logical inference, e.g., within the tool ManyOpt.

\section{CPS Framework} 
\label{section:framework}

We now introduce the NIST Framework for Cyber-Physical Systems, referred to as ``CPS Framework'' or simply ``Framework'' below. The Framework comprises a set of \emph{concerns and facets} related to the system under design or study. This section will clarify the intent and purpose of the framework, as well as its extensible and modifiable nature. The reader interested in documentation of the CPS Framework is directed to the three volume NIST \emph{Framework for Cyber-Physical Systems}:
\begin{itemize}
\item SP~1500-201 {\small (https://www.nist.gov/publications/framework-cyber-physical-systems-volume-1-overview)} \cite{CPSFramework-vol1}

\item SP~1500-202 {\small (https://www.nist.gov/publications/framework-cyber-physical-systems-volume-2-working-group-reports)} \cite{CPSFramework-vol2}

\item SP~1500-203 {\small (https://www.nist.gov/publications/framework-cyber-physical-systems-volume-3-timing-annex)} \cite{CPSFramework-vol3}
\end{itemize}

The CPS Framework provides the taxonomy and methodology for designing, building, and assuring cyber-physical systems that meet the expectations and concerns of system stakeholders, including engineers, users, and the community that benefits from the system's functions. The Framework comprises a set of concerns about systems, three development facets and a notion of functional decomposition suited to CPS. A CPS often delivers complex functions that are ultimately implemented in a multitude of collaborating systems and devices. This collaboration or interaction can occur through the exchange of information or the exchange of energy. We refer to the former as logical interaction and the latter as physical interaction. 

The functional decomposition of the Framework breaks a CPS down into functions or sets of functions, as follows:
\begin{itemize}
\item the \emph{Business Case}, a name and brief description of what the system is or does

\item the \emph{Use Case}, a set of scenarios or step-by-step description of ways of using the system and the functions that realize those steps

\item the \emph{Allocation of Function} to subsystems or actors~--~expressed in the terminology of Use Cases

\item the \emph{Physical-Logical Allocation}: allocation of given subsystem functions to physical or logical implementation.
\end{itemize}

As an example, consider a simplified version of an automated vehicle CPS for \emph{automated emergency braking}. The business case is a \emph{``vehicle system that detects objects and brings the vehicle safely to a stop without colliding with the obstacle.''} A corresponding use-case scenario consists of a \emph{sensor array} detecting an object and sending a \emph{braking torque request} to the braking system, where the amount of torque requested is based on a calculation of the distance to the object. The underlying subsystems or actors are the sensor array and the braking system which carries out the calculation and converts the request to an amount of electric power applied to components that produce the appropriate amount of hydraulic pressure on the braking calipers. The sensors are physical, the communication of the request is logical and the braking system is capable of both logical and physical function~--~it does calculation and creates hydraulic pressure. 

Next, we describe how the set of concerns of the CPS Framework is organized and \emph{applied to a function} in the functional decomposition of a CPS. The concerns of the Framework are represented in a multi-rooted, tree-like structure (a ``forest'' in graph theory), where branching corresponds to the \emph{decomposition of concerns}. We refer to this structure as the \emph{concern tree} of the CPS Framework. The concerns at the roots of this structure, the highest level concerns, are called \emph{aspects} and there are nine of them, one of which being \emph{Trustworthiness}.

A concern about a given system reflects consensus thinking about method or practice, involved in addressing the concern, and in some cases consensus-based standards describing that method or practice. This method or practice is applied to each function in the functional decomposition of the system and application of a concern to a function results in one or more properties to be required of that function in order to address the concern in question. A concern may be seen as a branch in the concern tree, consisting of the root name followed by a (possibly empty) sequence of concern element names in the branch, separated by periods or dots. In the Trustworthiness aspect, e.g., we have the concern $T\!rustworthiness.Security.C\!ybersecurity.C\!on\!f\!identiality$ that may be abbreviated as, e.g., $\confi$. A sample property, meant to address this concern about data exchanged between components of a system, is use of encryption of some kind (e.g. AES or DES). A property is 
appended to the concern tree branch in block parentheses. Here, 
$\confi[AES\!\!-\!\!encr]$ states that concern $\confi$ is intended to be addressed by the use of AES encryption. 

The facets of the CPS Framework are sets of activities, characteristic of a \emph{mode of thinking} about the development of a system. These facets are \emph{conceptualization}, \emph{realization}, and \emph{assurance}. We refer to the CPS Framework documentation noted above for their complete explanation. The output of the conceptualization facet is a \emph{Model of the CPS}, consisting of properties of the CPS with an indication of the concerns that gave rise to the properties. The output of the realization facet is the CPS itself. And, finally, the output of the assurance facet is an \emph{Assurance Case for each concern applied to the CPS}. The assurance case is sorted by the concerns applied to the CPS and consists of \emph{assurance judgment(s)}, comprised of:
\begin{itemize}
\item \emph{Properties} of the CPS and the concerns that resulted in the addition of those properties to the Model of the CPS.

\item \emph{Argumentation:} consensus or authority-based description of criteria for concluding that a property, intended to address a concern, has been established of the CPS.

\item \emph{Evidence:} information, accessible to stakeholders, that the criteria used in this argumentation are indeed met.

\item \emph{Uncertainty:} qualitative or quantitative representation of the uncertainty associated with the evidence that the criteria are met.
\end{itemize}

\subsection{Why a CPS Framework?}
There are many critical concerns about the CPSs that surround us or that we depend upon~--~including the sub-concerns of $T\!rustworthiness$: $Sa\!f\!ety$, $Security$, $Privacy$, $Resilience$, and $Reliability$. The urgency of addressing such concerns has only increased with the rapid deployment of CPS in domains such as transportation, medical care, and energy. There are clear needs to design for trustworthiness and monitor the \emph{trustworthiness status} of these CPS, since components can fail and new threats can emerge over time.

\begin{figure}
\begin{center}
\includegraphics[width=\textwidth]{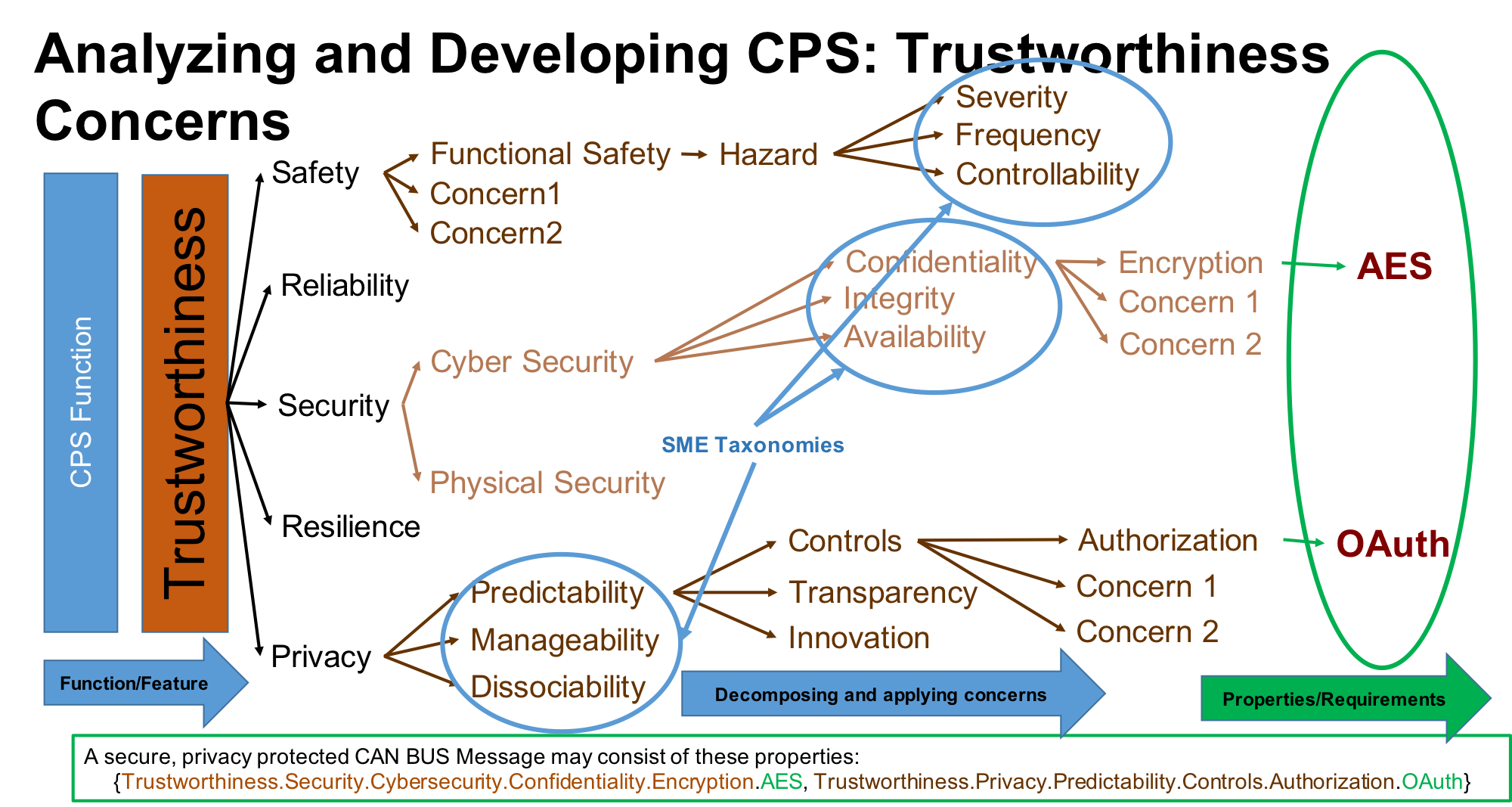}
\end{center}
\caption{Decomposition tree for Trustworthiness Concerns\label{fig:trustworthiness}}
\end{figure}

The CPS Framework provides a \emph{CPS Normal Form}: any CPS can be analyzed through the same analytical lenses of the CPS Framework (see, e.g., Figure~\ref{fig:trustworthiness}), resulting in the functional decomposition of the CPS annotated with its concerns and the properties introduced in its Model in order to address those concerns. Given two CPS and their respective analyses, we may thus compare these CPS directly, one concern at a time.

Subsequent to the initial release of the CPS Framework, referenced above, NIST modeled the CPS Framework using the Unified Modeling Language (UML) and generated an XML schema or type structure of the CPS Framework. This effort was labeled the \emph{CPS Framework Open Source Project} and, as a follow up, NIST held a CPS Framework Open Source Workshop on September 19, 2017. The intent of this modeling effort, in the format of XML, was to:
\begin{itemize}
\item Represent CPS in a common data exchange format (to facilitate concern-focused design collaboration)

\item Provide an IT-based mechanism for comparing the concern-integrity of CPS (to enable a concern-centric assessment of CPS composition)

\item Facilitate a concern-focused interface to CPS (to assess and monitor the status of a CPS relative to measurable properties and their associated concerns).
\end{itemize}

\begin{figure}
\begin{center}
\includegraphics[width=\textwidth]{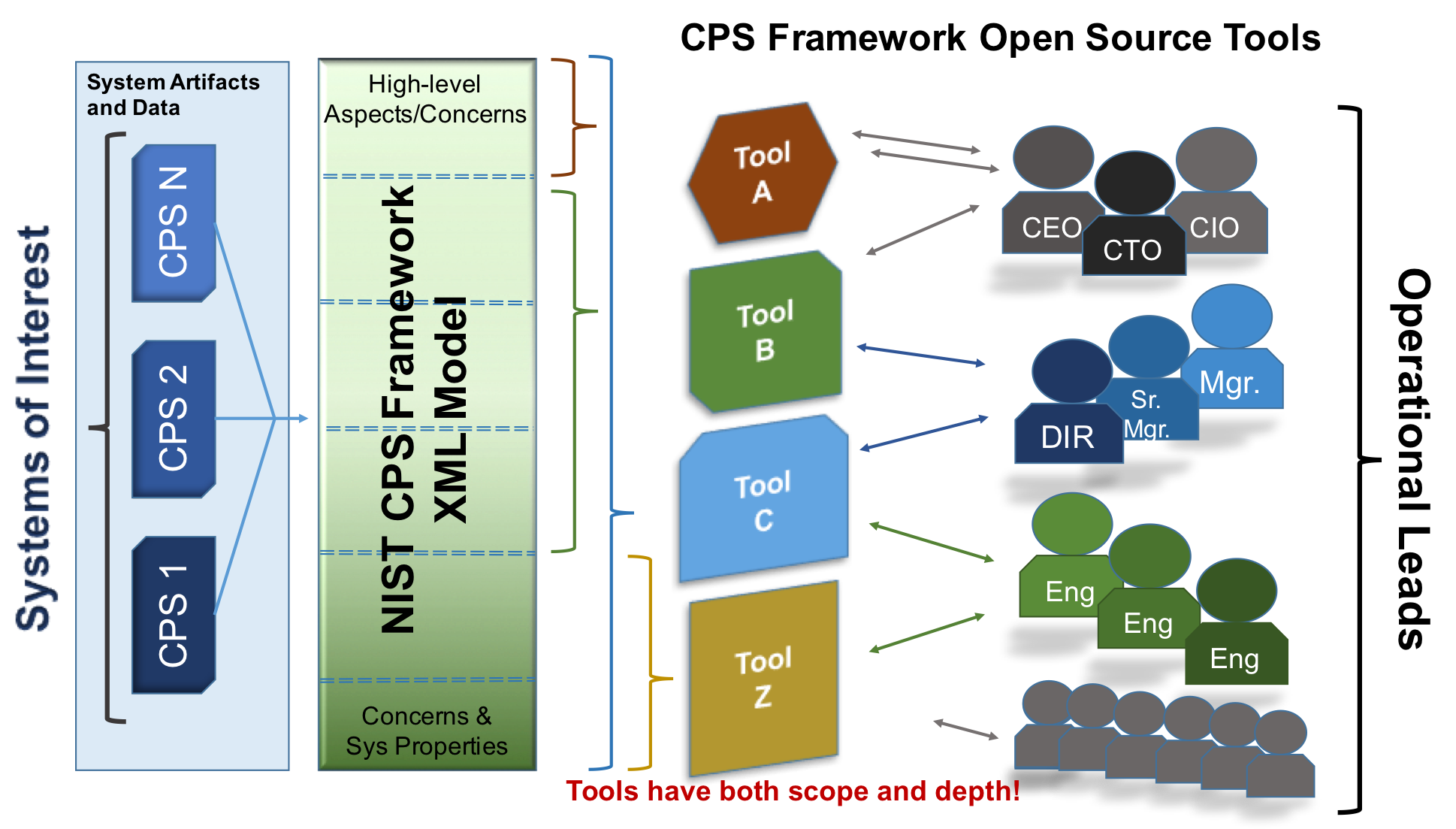}
\end{center}
\caption{Open-source tools supporting the CPS Framework\label{fig:tools}}
\end{figure}

This CPS Framework and the Open Source technology, depicted in Figure~\ref{fig:tools}, are essential to understanding critical performances of CPSs incrementally, 
from the perspective of CPS development, deployment, and adoption.

\subsection{Relation of this Paper to the CPS Framework}
The work presented in this paper is an extension of the open source project reported above. The UML/XML modeling provides a concern-focused portal to CPS. It demonstrates a methodology to reason about mutual dependencies and conflicts in requirements that need to be taken into consideration during the design, deployment, and operational stages. 
Information  
needed for such reasoning, can be manually entered or obtained from a continuous feed from a sensor array designed to measure base requirement satisfaction. It can also be generated in other ways, depending on the nature of the system. 
The reasoning engine described in this paper is realized by modeling a CPS through  ontologies based on the CPS Framework.

The semantic relationship of the CPS framework to the work reported here is as follows. The above synopsis of pertinent CPS Framework concepts and approaches featured the forest of concerns, where each tree represents an aspect. There are two types of nodes, concern elements and property nodes, as well as two types of edges: those that represent decomposition of concerns and those that connect concern elements to properties. Both types of edges should be thought of as \emph{AND edges}, meaning that the satisfaction of the parent concern requires that all of the children nodes be satisfied. In our approach, we address a concern by satisfying its node in the concern tree. This means that a concern element satisfies all its children~--~which are refined concerns, and that a property node satisfies all of its properties. Logical conjunction is therefore the basis of this satisfaction relation.

\section{A CPS Framework Ontology} 

\label{section:ontologybackground}

At the core of this approach is an ontology of the CPS Framework and of a CPS of interest. An ontology is a formal, logic-based representation that supports reasoning by means of logical inference. In this paper, we adopt a rather broad view of this term: by ontology, we mean a collection of statements in a logical language that represent a given domain in terms of \emph{classes} (i.e., sets) of objects, \emph{individuals} (i.e., specific objects), relationships between objects and/or classes, and logical statements over these relationships.

In the context of the trustworthiness of CPS, for instance, an ontology might define the high-level concept of ``Concern'' with its refinement of ``Aspect.'' All of these will be formalized as classes and, for Aspect, subclasses. Specific concerns will be represented as individuals: $T\!rustworthiness$ as an individual of class Aspect, $Security$ and $Cybersecurity$ of class Concern.
Additionally, a relation ``has-subconcern'' might be used to associate a concern with its sub-concerns. Thus, Aspect ``has-subconcern'' $Security$, which in turn ``has-subconcern'' $Cybersecurity$. By introducing a property ``satisfied,'' one could also indicate which concerns are satisfied. 

Inference can then be applied to propagate ``satisfied'' and other relevant properties and relations throughout the ontology. For example, given a concern that is not ``satisfied,'' one can leverage relation ``has-subconcern'' to identify the concerns that are not satisfied, either directly or indirectly, because of it. 

In practice, it is often convenient to distinguish between the factual part, $\Omega$, of the ontology (later, simply called ``ontology''), which encodes the factual information (e.g., $T\!rustworthiness$ ``has-subconcern'' $Security$), and the \emph{axioms}, $\Lambda$, expressing deeper, often causal, links between relations (e.g., a concern is not satisfied if any of its sub-concerns is not satisfied). Further, when discussing reasoning tasks, we will also indicate, separately, the set $\mathcal{Q}$ of axioms encoding a specific reasoning task or query.

\section{Applying Ontology and Reasoning to CPS}
\label{section:reasoningbackground}
By leveraging a logic-based representation of a domain of interest, one can apply inference and draw new and useful conclusions in a principled, rigorous way. In essence,  our approach is agnostic to any specific choice of logical language and inference mechanisms. Axioms expressed in the used logical language formalize the queries one is interested in answering, the type of reasoning that can be carried out, and any additional contextual information.
Thus, given an ontology $\Omega$, a set of axioms $\Lambda$, and an inference relation $\entails$, we say that $\Delta$ is an answer to the (implicit) query iff
\[
\Omega \cup \Lambda \entails \Delta.
\]
where $\cup$ denotes the union of two sets. For instance, in the language of propositional logic, given knowledge that some proposition $p$ is true and that $p$ implies some other proposition $q$, one can infer that $q$ is also true, i.e.:
\[
\{p, p \supset q\} \entails \{q\}.
\]
In the context of cybersecurity, $p$ might be true when a cyberattack has occurred and $p \supset q$ might formalize an expert's knowledge that, whenever that cyberattack occurs, a certain system becomes inoperative (proposition $q$). The logical inference represented by symbol $\entails$ allows to draw the conclusion that, as a result of the cyberattack, the system is now inoperative. For increased flexibility of representation, we use here a non-monotonic extension of propositional logic, called Answer Set Programming (ASP) \cite{gl91,mt99,bar03}. ASP is a rule-based language, where a rule is a statement of the form
\begin{equation}\label{eq:rule}
h_1 \lor h_2 \lor \ldots h_k \hif l_1, \ldots, \l_m, \lpnot l_{m+1}, \ldots, \lpnot l_n.
\end{equation}
Every $h_i$ and $l_i$ is a \emph{literal}, i.e. an atomic proposition analogous to $p$ and $q$ above, optionally prefixed by the negation symbol $\neg$ to express its negation. Intuitively, Equation~(\ref{eq:rule}), hereafter referred to as~(\ref{eq:rule}), states that, if $l_1, \ldots, l_m$ hold and there is no reason to believe (the {\tt not} keyword in~(\ref{eq:rule})) that $l_{m+1}, \ldots, l_n$ hold , then one of $h_1, \ldots, h_k$ must hold. Thus, the ASP counterpart of the propositional logic implication $p \supset q$ is $q \hif p$. Suppose proposition $r$ represents the fact that the system is patched against the cyberattack. To make conservative predictions about the system state after a cyberattack, we might want to conclude that the system should be expected to be inoperative unless there is positive evidence that it was patched. This can be represented in ASP by:
\[
q \hif p, \lpnot r.
\]
Note the difference between $\neg r$ and $\lpnot r$. The former is true if we have explicit evidence that the system has not been patched. The latter does hold whenever we have that explicit evidence, but also whenever we simply do not know if it was patched or not. Depending on specific needs, Answer Set Programming allows either type of expression. (This type of default reasoning is an example of the greater flexibility of representation that motivates our use of ASP in this paper.)

Although ASP is propositional in nature, we follow common representational practice and allow for a literal to include a list of arguments, possibly comprising logical variables. For example, we may write $q(s_1)$ to indicate that it is system $s_1$ that is inoperative. Similarly, given a variable $\mathtt{X}$, we may use
\[
q(\mathtt{X}) \hif p, \lpnot r(\mathtt{X}).
\]

\noindent to say that any system $\mathtt{X}$ that is not known to be patched should be assumed to have been made inoperative by the cyberattack.

\subsection{Naming Conventions} 
The decomposition of a CPS identifies resources that may satisfy properties. Suppose that $cam$ is a camera, a subsystem of an autonomous car, and that $mem$ is a memory sub-system of $cam$; we will examine this system in more detail later. Then $cam\_mem[encr]$, e.g., is a Boolean predicate that is true if the memory $mem$ of camera $cam$ uses encryption. Properties thus have form $SystemPath[prop]$ where $SystemPath$ identifies a system component or part, with subcomponents indicated by the underscore symbol, and $prop$ a property that this part may enjoy. We interpret two such properties to be equal only if their actual names are equal: $cam\_mem[encr]$ and $cam\_mem'[encr]$, e.g., are different properties as the same encryption is applied to different memories of the same camera $cam$. Properties $SystemPath[prop]$ also have a semantic context $ConcernPath$ that articulates which (sub)concern of an aspect this property is trying to address. Property $cam\_mem[encr]$, e.g., may have context 
$T\!r\!ustworthiness.Security.C\!ybersecurity.C\!on\!f\!identiality$, where we use the dot operator ``.'' in $ConcernPath$ to distinguish this easily from navigations in $SystemPath$. In our semantics below, a property may be either true or false (i.e., satisfied or non-satisfied). These truth values in turn influence the satisfaction of concerns and aspects.
Below, we elide
details of such context or of system paths; e.g., $\confi$ may abbreviate $T\!r\!ustworthiness.Security.C\!ybersecurity.C\!on\!f\!identiality$.

\subsection{Formalization} 
\label{section:formalization}

For sake of illustration, we consider a lane keeping/assist (LKAS) use case centered around an advanced car that uses a camera and a situational awareness module (SAM) for lane keeping/assist. The SAM processes the video stream from the camera and  controls, through a physical output, the automated navigation system. The camera and the SAM may use encrypted memory and secure boot. Safety mechanisms in the navigation system cause it to shut down if issues are detected in the input received from the SAM. 
This use case is chosen because it encompasses major component types of a CPS, and lends itself to various non-trivial investigations. Through this use case, we will highlight the interplay among trustworthiness concerns, as well as their ramifications on other CPS aspects, such as the functional aspect. 

For sake of presentational simplicity, we will assume that the camera is capable of two recording modes, one at 25 fps (frames per second) and the other at 50 fps. The selection of the recording mode is made by the SAM, by acting on a flag of the camera's configuration. 
It is assumed that two camera models exist, a basic one and an advanced one. Either type of camera can be used when realizing the CPS. Due to assumed technical limitations, the basic camera is likely to drop frames if it attempts to record at 50 fps while using encrypted memory.

In our approach, the formalization of a CPS is organized along multiple levels: (L1) aspects and concerns; (L2) properties; (L3) CPS configuration; (L4) actions; (L5) constraints, dependencies and trade-offs; and (L6) satisfaction axioms. Level L1 and L6 form the \emph{CPS-independent specification}, since aspects and concerns are independent of the specific CPS being modeled. Levels L2-L5 comprise the \emph{CPS-dependent specification}, as the information included in them depends on the CPS being modeled. Furthermore, levels L1 and L2 formalize the concepts from the definition of the CPS Framework. Levels L3-L5 extend the CPS Framework in order to provide details needed for reasoning about the behavior of a CPS of interest. Level L6 provides the semantics of the formalization. Next, we describe our approach through its application on the LKAS use case.

\textbf{Formalization of aspects and concerns.} The formalization of aspects and concerns is shared by all CPSs. The nodes of a concern tree are represented by individuals of class \emph{Concern}.  The root nodes of the concern trees are a particular kind of concern, and so they are placed in a class (\emph{Aspect}) that is a subclass \emph{Concern}. Following the definition of the CPS\ Framework, class \emph{Aspect} includes individuals \emph{Trustworthiness}, \emph{Timing} and \emph{Functional} for the corresponding aspects, while class \emph{Concern} includes individuals \emph{Security}, \emph{Cybersecurity}, \emph{Functionality}, etc.

Edges linking aspects and concerns are represented by the relation $\subcon$, which is a representation of ``sub-concern.'' Thus, an edge from a concern $x$ to a concern $y$ is formalized by a statement $\subcon(x,y)$. Statement 
$$\subcon(T\!r\!ustworthiness,Security)$$

\noindent e.g., formalizes that the Security concern is a direct sub-concern of the Trustworthiness aspect in our
LKAS use case.
Concerns $Cybersecurity$ and $\confi$ are linked similarly.
\begin{figure}
\includegraphics[clip=true,trim=120 60 120 0,width=1\textwidth]{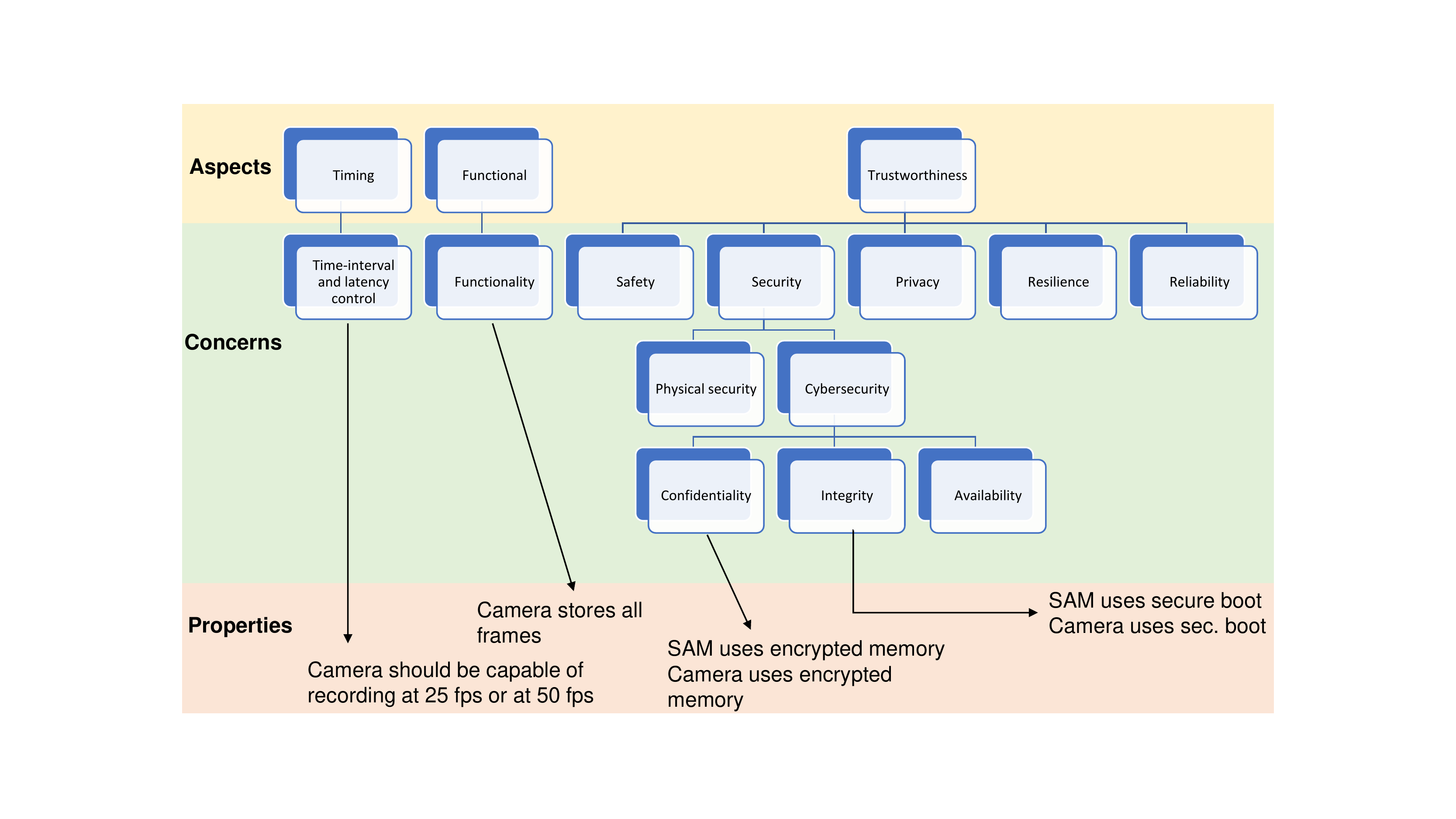}
\caption{LKAS use case: pertinent part of the concern forest
}
\label{fig:aspects-concerns}
\end{figure}

\textbf{Formalization of properties.} Properties of a CPS are represented by individuals of class \emph{Property}. An edge that links a property with an aspect or concern is represented by relation $\addby$, which stands for ``addressed by.''
Let us suppose that, in the LKAS use case, both SAM and camera must use encrypted memory for the confidentiality concern to be satisfied (see Figure \ref{fig:aspects-concerns}). We may express this by
two statements $\addby(\confi,SAM\_mem[encr])$ and $\addby(\confi,cam\_mem[encr])$.
Similarly, the fact that SAM and camera must use secure boot for the integrity concern to be satisfied is expressed by the statements $\addby(Integrity,$ $SAM\_boot[sec])$ and $\addby(Integrity,cam\_boot[sec])$. 

Another property, referred to below, is $cam[\conrate]$, stating that camera $cam$ stores all frames, i.e. does not drop any frames. Note that, in the LKAS use case, the car heavily depends on the camera for proper lane keeping/assist: not dropping any frames is essential for satisfaction of the functionality concern.

\textbf{Formalization of configurations.} Properties do not necessarily capture all possible configurable features of a CPS, but only those on which concerns are defined. For instance, in the LKAS use case, there is a choice between using the basic camera or the advanced camera. We describe the choice between the two as part of the configuration of the CPS.  Thus, the formalization includes a class \emph{Configuration}. Each individual of this class represents a different configuration feature, e.g. $cam[\basicone]$ is used for the selection of a type of camera $cam$. Similarly to properties, configurations can be true or false in a given state of the CPS. In fact, their truth value is essential in defining the configuration of the CPS for a scenario of interest. Truth values of properties and configurations are specified by relation \emph{obs}, where a statement $obs(x,true)$ declares that property or configuration $x$ is (observed to be) true. Observability of falsity is represented in a similar way.

\textbf{Formalization of actions.} We use the term ``action'' to denote both those actions that are within the control of an agent (e.g., actions a driver may take), and those actions that occur spontaneously, e.g.\ triggered by a particular state of the CPS such as the automatic disabling of the LKAS capability if the camera malfunctions. 
The formalization includes a suitable class \emph{Action}
and individuals for the actions of interest. In the LKAS use case, we consider the occurrence of a cyberattack, and formalize it by means of the individual/action
labeled \emph{Attack}. The case in which the automated navigation system shuts down is modeled by an individual $NavShutdown$. When the configuration of a CPS can be modified at run-time, suitable actions $MakeTrue(c)$ and $MakeFalse(c)$ may also be introduced, where $c$ is the configuration the action affects. For example, in the LKAS use case, we consider actions $MakeTrue(cam[\basicone])$ and $MakeFalse(cam[\basicone])$, which, respectively, switch on or switch off the basic camera.

\textbf{Formalization of constraints, dependencies, trade-offs.} An additional feature of our model is the ability to establish causal links between concerns, properties, configurations,  and actions. This is accomplished by the reasoning over statements. Table~\ref{tab:constraints} lists types of statements, their syntactic expressions as judgments, and their corresponding encodings for the ASP reasoner. The logical encodings of the statements are used to implement reasoning capabilities discussed later in the paper. 
\begin{table}[htbp]
\begin{small}
\begin{center}
\begin{tabular}{|l|l|l|}
\hline
\textbf{Statement type} & \textbf{Syntax} & \textbf{Encoding for reasoner} \\
\hline
\specialcell[c]{\hspace*{-.2in}$\mbox{Property}$\\$\mbox{dependency}$} & 
\specialcell[c]{$\Gamma \impactspos \pi$ \\
                    $\Gamma \impactsneg \pi$} &
\specialcell[c]{\!\!$impacted(pos/neg,\pi,\mathtt{S}) \hif$ \\$holds(\Gamma,\mathcal{S})$}\\
\hline
\specialcell[c]{$\mbox{Default property}$\\$\mbox{\hspace*{-.56in}value}$}
& 
\specialcell[c]{$\sigma\ \defaults\ true$ \\
$\sigma\ \defaults\ f\!alse$} &
\!\!$\defaults(\sigma,true/false)$\\
\hline
Effects of actions & $a \causes \pi \lawif \Gamma$&
\specialcell[c]{\hspace*{-.5in}$holds(\pi,\mathcal{S}+1) \hif$ \\$\ \ holds(\Gamma,\mathcal{S}), occurs(a,\mathtt{S})$}\\
\hline
Triggered actions & $\Gamma \triggers a$ &
\!\!$occurs(a,\mathtt{S}) \hif holds(\Gamma,\mathtt{S})$\\
\hline
\end{tabular}
\end{center}
\end{small}
\vspace*{-.2in}  
\caption{Constraints, dependencies, and trade-offs where $\Gamma$, $\pi$ range over (sets of) propositions and $a$ over actions}\label{tab:constraints}
\vspace*{.04in}  
\end{table}
For an example of a property dependency statement, recall that the use of encrypted memory causes the basic camera to drop frames if it attempts to record at 50 fps. We formalize this by:
\begin{eqnarray}
cam\_mem[encr] \land \neg cam[\lowerrate] \land\ cam[basicOne] & {}\nonumber\\
\qquad \impactsneg  cam[\conrate] & {} \label{eq:impacts-neg} 
\end{eqnarray}

The statement states that, under the conditions specified, the $\conrate$ property is \emph{impacted negatively}, that is, is made false. If a property is impacted positively, $\impactspos\!$ is used instead. As shown in this example, properties and configurations can be negated by prefixing them by $\neg$. 
Let us list relevant aspects of concerns from the contexts of these properties: $\confi$ for $encr$, $Timing$ for $\lowerrate$, $Con\!f\!iguration$ for $\basicone$, and $F\!unctionality$ for $\conrate$.
In the case of $\conrate$, one may also want to specify that the property should be assumed to hold true in the absence of contrary evidence. This can be achieved by a statement:
\[
\conrate\ \defaults\ true
\]

\noindent The effects of actions on properties are given by statements borrowed from action language $\mathcal{AL}$ \cite{gelfond-1}, which has been designed specifically for a compact specification of the causal dependencies in complex domains.\footnote{While we find $\mathcal{AL}$ convenient, our approach does not depend on a particular choice of language. Other languages, e.g. PDDL, can be easily incorporated into our approach.} Let us say, for instance, that in the LKAS use case a cyberattack may force the camera to record at 50 fps. Using action \emph{Attack}, introduced earlier, this may be formalized by a law \[
Attack \causes \neg \lowerrate.
\]
The last type of statement from Table \ref{tab:constraints} describes the spontaneous triggering of actions when suitable conditions are satisfied. To illustrate this, recall that, in the LKAS use case, safety mechanisms in the navigation system cause the navigational system to shut down if issues are detected in the input received from the SAM. One obvious circumstance in which this will happen is if the system is not fully functional. This link can be formalized by the trigger:
\begin{equation}\label{eq:navshutdown}
\neg Functional \triggers NavShutdown.
\end{equation}

\textbf{Axioms.} Recall that our approach reduces the task of answering a query of interest to that of finding one or more answers, $\Delta$, such that $\Omega \cup \Lambda \entails \Delta$ holds, where the ontology $\Omega$ and any supporting axioms $\Lambda$ are expressed in a logical language for the reasoner of choice~--~ASP in this paper. The statements presented so far can be easily translated into logic statements as seen in the last column of Table \ref{tab:constraints}, e.g. (\ref{eq:impacts-neg}) translates to
\begin{equation}\label{eq:impacted}
\begin{array}{l}
impacted(neg,cam[\conrate],\mathtt{S}) \hif \\
\aspindent holds(cam\_mem[encr],\mathtt{S}), \\
\aspindent \neg holds(cam[\lowerrate],\mathtt{S}), \\
\aspindent holds(cam[basicOne],\mathtt{S}).
\end{array}
\end{equation}

\noindent where $holds$ is an auxiliary relation that states that its argument holds at a discrete step $\mathtt{S}$ in the evolution of the CPS. As we will demonstrate later, the inclusion of a step argument makes it possible to 
analyze the evolution of the CPS over time in response to possible events.

It remains to formalize
the meaning of relation $impacted$ in terms of the effect on the truth value of $cam[\conrate]$. In our approach, this is accomplished by a set of axioms that complete the translation of the statements from Table \ref{tab:constraints} and, additionally, enable reasoning about the satisfaction of properties, concerns, and aspects. Due to space considerations, we focus the presentation on the latter, shown in Figure \ref{fig:satisfaction-axioms}.
\begin{figure}[htbp]
\centering
\begin{equation}\label{eq:sat-concern}
\begin{array}{rl}
\neg holds(sat(\mathtt{C}),\mathtt{S}) \hif  \!\!\!\! &
\addby(\mathtt{C},\pi),\\
&\lpnot holds(\pi,\mathtt{S}).\\
\end{array}
\end{equation}
\vspace*{-.15in}
\begin{equation}\label{eq:sat-recursive}
\begin{array}{rl}
\neg holds(sat(\mathtt{C_1}),\mathtt{S}) \hif \!\!\!\!&
\subcon(\mathtt{C_1},\mathtt{C_2}), \\
&\neg holds(sat(\mathtt{C_2}),\mathtt{S}).
\end{array}
\end{equation}
\vspace*{-.09in}
\begin{equation}\label{eq:defaults}
\begin{array}{rl}
holds(\mathtt{X},\mathtt{S}) \hif  \!\!\!\!&
\defaults(\mathtt{X},true),\\
&\lpnot \neg holds(\mathtt{X},\mathtt{S}).
\end{array}
\end{equation}
\vspace*{-.07in}
\begin{equation}\label{eq:obs-holds-a}
\begin{array}{l}
holds(\pi,0) \hif obs(\pi,true). \\
\end{array}
\end{equation}
\begin{equation}\label{eq:obs-holds-b}
\begin{array}{l}
\neg holds(\pi,0) \hif obs(\pi,false). \\
\end{array}
\end{equation}
\caption{Satisfaction-related axioms for LKAS use case}\label{fig:satisfaction-axioms}
\end{figure}

Axiom (\ref{eq:sat-concern}) intuitively states that a concern is not satisfied if any of the properties that address it does not hold. This ensures that the lack of satisfaction of a property $\pi$ is propagated to the concern(s) that are addressed by $\pi$ according to the $\addby$ statements provided by the formalization of properties. The lack of satisfaction is then propagated up the relevant concern tree by axiom (\ref{eq:sat-recursive}) according to the concern-concern dependencies specified by the $\subcon$ statements in our ontology.

One may note that axioms (\ref{eq:sat-concern})-(\ref{eq:sat-recursive}) only address the lack of satisfaction of properties and concerns. The specification of the notion of satisfaction is completed by $\defaults$ statements saying that all properties and concerns are satisfied by default, by axiom (\ref{eq:defaults}), which embodies the semantics of the $\defaults$ statements, and by axioms (\ref{eq:obs-holds-a})-(\ref{eq:obs-holds-b}), which link the observations about the initial state to auxiliary relation $holds$.

Thus, if the basic camera is used with encrypted memory while recording at 50 fps, (\ref{eq:impacted}) makes it possible to conclude that property $\conrate$ is not satisfied. In turn, (\ref{eq:sat-concern}) yields that \emph{F\!unctionality} is not satisfied. Finally, (\ref{eq:sat-recursive}) concludes that the functional aspect is not satisfied.

\subsection{Reasoning} 
\label{section:reasoning}
The formalization presented above makes it possible to reason about aspects and concerns of a CPS, their interdependencies, and their implications in relation to the other systems the CPS may interact with. Now, we illustrate these reasoning capabilities by focusing 
mostly on the trustworthiness concerns, but the reasoning mechanisms we established can be applied to arbitrary parts of the aspects hierarchy.

\textbf{Concern tree.} For the LKAS CPS, let the basic camera be used, SAM and camera use encrypted memory and secure boot, and the recording rate be set to 50 fps. Once aspects, concerns, properties, and configurations are formalized as described earlier, this system state is formalized by the statements:
\[
\begin{array}{l}
obs(\basicone,true),\ 
obs(cam\_mem[encr],true), \\
obs(cam\_boot[sec],true),\ 
obs(cam[\lowerrate],false), \\
obs(SAM\_mem[encr],true),\ 
obs(SAM\_boot[sec],true) \\
\end{array}
\]

\noindent By inspecting Figure \ref{fig:aspects-concerns}, one can see that the confidentiality concern is satisfied. From a technical perspective a query ``is $\chi$ satisfied by the design of the CPS?'', where $X$ is a property (e.g., $\conrate$) or concern, is answered by checking whether
$\Omega \cup \Lambda \entails holds(\chi,0).$
By specifying a different time step, one can also check whether the query is satisfied at run-time. In our running example, starting from the observation that encrypted memory is used, axiom (\ref{eq:sat-concern}) allows one to conclude that $\Omega \cup \Lambda \entails holds(sat(\confi),0)$. Similarly, one can formally conclude $holds(sat(Integrity),0)$. From (\ref{eq:sat-recursive}) and (\ref{eq:defaults}), it also follows that \emph{Cybersecurity} is satisfied and, in turn, all concerns up to \emph{Trustworthiness}. Thus the LKAS CPS is deemed to be trustworthy.

On the other hand, $\Omega \cup \Lambda$ entails that both statements $\neg holds(\conrate,0)$ and $\neg holds(sat(Functional),0)$ are true and, recursively, the \emph{Functionality} concern and the \emph{Functional} aspect are thus not satisfied.

\textbf{All-sat.}
One may also want to check whether all aspects are satisfied. This query is encoded by the set $\mathcal{Q}$ of axioms: 
\begin{equation}\label{eq:sat-all}
\!\!\!\begin{array}{l}
sat(all)\ \defaults\ true. \\
\neg holds(sat(all),\mathtt{S}) \hif
aspect(\mathtt{A}),\ \neg holds(sat(\mathtt{A}),\mathtt{S}).
\end{array}
\end{equation}
These axioms introduce a ``meta-aspect'' $all$, representing the satisfaction of the entire concern forest, and state that it is enough for one aspect not to be satisfied, to cause the concern forest not to be satisfied as a whole. In our example, one can check that $\Omega \cup \Lambda \cup \mathcal{Q} \entails \neg holds(sat(all),0)$. In fact, as we saw in the previous paragraph, $\neg holds(sat(Functionality),0)$ is entailed. This is sufficient to trigger (\ref{eq:sat-all}) and derive $\neg holds(sat(all),0)$. That is, the CPS is deemed to be trustworthy, but does not satisfy the functional aspect: therefore, the concern forest, as a whole, is not satisfied.

\textbf{Partial synthesis/Design completion.} Our approach also allows for the completion of a partially specified CPS design so that desired constraints are satisfied. Let $\gamma$ be the requirement that must be satisfied, e.g. $sat(\confi)$ or $sat(all)$. The corresponding query is encoded by the set $\mathcal{Q}$ of axioms:
\[
\begin{array}{l}
holds(\pi,0) \lor \neg holds(\pi,0). \\
\bot \hif \lpnot holds(\gamma,0).
\end{array}
\]
where the first rule states that any property $\pi$ can be true or false\footnote{Note that the axioms of $\Lambda$ prevent the selection of truth values that conflict with $obs(\cdot,\cdot)$ statements provided.} and the second says that $holds(\gamma,0)$ must be true in every solution/answer returned. For illustration, let us complete the partial design:
\[
\begin{array}{l}
obs(\basicone,true),\ 
obs(cam\_boot[sec],true), \\
obs(cam[\lowerrate],f\!alse),\ 
obs(SAM\_mem[encr],true), \\
obs(SAM\_boot[sec],true).
\end{array}
\]
Note that the design does not specify whether the camera uses encrypted memory or not. Let us suppose that we are interested in finding a completion of the design in which the LKAS CPS is trustworthy. To do that, we specify $\gamma$ to be $sat(Trustworthiness)$. One can now check that $\Omega \cup \Lambda \cup \mathcal{Q}$ entails\footnote{To be precise, credulous entailment is used in this example.} $holds(cam\_mem[encr],0)$. In fact, the completion of the design in which the camera uses encrypted memory makes the CPS trustworthy for purposes of the design analysis.

\textbf{What-if.} A \emph{What-if} reasoning task studies how the CPS is affected by the occurrence of actions, in terms of which properties hold, which concerns are satisfied, and which other actions may be triggered. Let the expression $occurs(a,s)$ denote the occurrence of action $a$ at step $s$ and let a history $\mathcal{H}$ be a set of such expressions. A query ``is $\chi$ satisfied at step $s'$?'', where $\chi$ is a property (e.g., $\conrate$) or concern and $s'$ is a step during or after history $\mathcal{H}$, is answered by checking whether
$\Omega \cup \Lambda \cup \mathcal{H} \entails holds(\chi,s').$

A query ``does action $a$ occur at step $s'$?'' is answered by checking whether
$\Omega \cup \Lambda \cup \mathcal{H} \entails occurs(a,s').$
Obviously, the same mechanism allows for answering more general questions, such as ``is $X$ satisfied (or not satisfied) at some point during $\mathcal{H}$?'' and ``which actions are triggered during $\mathcal{H}$?''.
In reference to the LKAS use case, let us consider a scenario in which, initially, the basic camera is used, SAM and camera use encrypted memory and secure boot, and the recording rate is set to 25 fps. Clearly, the functional aspect is satisfied by the CPS. We want to study whether the functional aspect  remains satisfied after $occurs(Attack,0)$. That is, we need to check whether
\[
\Omega \cup \Lambda \cup \mathcal{H} \entails holds(sat(F\!unctional),1).
\]
Note the use of step $1$ in the query, which corresponds to the step that follows the hypothesized occurrence of $Attack$. One can check that the answer to the query is negative. In fact, as we discussed earlier, the attack forces the camera to record at 25 fps. From (\ref{eq:impacted}), it follows that the camera will begin to drop frames, which in turn affects the functional aspect negatively. One may wonder whether there are any further side-effects -- for instance, whether any follow-up actions are triggered. This can be accomplished by checking if there is any other action $a$ that occurs at step $1$. Given that the functional aspect is no longer satisfied, (\ref{eq:navshutdown}) will cause $\Omega \cup \Lambda \cup \mathcal{H}$ to entail $occurs(NavShutdown,1)$, indicating that the navigation system will shut down. (Recall that $occurs(\cdot,\cdot)$ is derived from the $\triggers$ statement, as seen in Table \ref{tab:constraints}.)

\textbf{Mitigation.} The last reasoning task we illustrate is aimed at determining how the effects of a history can be mitigated. As before, let $\mathcal{H}$ be a set of occurrences of actions. We are interested in answering the query ``which mitigation measure can restore $\gamma$?'' where $\gamma$ is a concern or the meta-aspect $all$.\footnote{For illustration purposes, we focus on after-the-fact mitigation. It is not difficult to extend the technique to cover preventive measures.} To simplify the presentation, let us focus on the case in which all mitigation actions are executed concurrently after the last action of $\mathcal{H}$. Let $s^\#$ denote the corresponding step. The set $\mathcal{Q}$ of axioms that encode the query includes a rule of the form
$occurs(a,s^\#) \lor \neg occurs(a,s^\#)$
for every action $a$ that one is interested in allowing, as well as a rule
\[
\begin{array}{l}
\bot \hif \lpnot holds(sat(\gamma),s^\#\!+\!1).
\end{array}
\]
stating that it is impossible for $\gamma$ not to be satisfied. The question is answered by finding the set of actions $a$ such that $\Omega \cup \Lambda \cup \mathcal{H} \cup \mathcal{Q} \entails occurs(a,s^\#)$. In the LKAS use case, it is not difficult to check that the mitigation action returned by this process is 
$MakeFalse(cam[\basicone])$, indicating that the basic camera should be replaced by the advanced camera in order to compensate for the fact that the cyberattack is forcing the CPS to record at 50 fps.

If the underlying inference mechanism allows for finding multiple solutions, one can also use our approach to  find optimal solutions. For instance, one might ask ``which mitigation measures can restore $\gamma$ and involve the smallest number of actions?''. If ASP is the underlying logical formalism, the query can be easily encoded by extending $\mathcal{Q}$ by a rule:
\begin{equation}\label{eq:minimize}
<\!\sim occurs(\mathtt{A},s^\#).
\end{equation}
where ``$<\!\sim$'' is the advanced \emph{weak constraint} connective, requesting the minimization of occurrences of its right-hand side in any solution found.\footnote{It is possible to use other types of minimization as well.} 

To illustrate the task, consider a variation of the LKAS use case in which a SAM that is affected by the cyberattack can be patched (action \emph{Patch}) to force it to request 25 fps recording at all times. Let $\Omega$ and $\Lambda$ be modified accordingly and $\mathcal{Q}$ be expanded as described above. One can check that $\Omega \cup \Lambda \cup \mathcal{H} \cup \mathcal{Q}$ now entails two alternative solutions: $occurs(Patch,s^\#)$ and $occurs(MakeFalse(cam[\basicone],s^\#)$ . While, in principle, another possible mitigation consists in \emph{both} replacing the basic camera \emph{and} patching the SAM, it is ruled out by (\ref{eq:minimize}) because it is non-minimal.

\section{Discussion}\label{section:discussion} 
Kolbe et al.\ \cite{Kolbe2017} stress the importance of situational awareness in complex systems and the  benefits of ontologies to enable a rich context that permits the developers and operators to model a large number of situations.  Others, e.g. Gyrard et al.\ \cite{Gyrard2015}, stress the advantages of using ontologies and logical reasoning for cross-domain application development. Our experimental use cases illustrate that the richer context brought forward by the proposed approach supports more holistic insights into complex systems, their development, and operations, and allows developers to model rich contexts and anticipate issues, constraints, and conflicts that are not self-evident and are multi-domain in nature. 

Cyber-physical contexts are very diverse and have diverging operational and design requirements. Different emphasis is needed to design safe and secure aircraft, a smart meter, a connected medical device, or a connected home appliance. Nevertheless, similar technologies and fundamental design principles are used to build these differing systems, and they share dependencies on similar or connected infrastructure technologies. When specific technologies are analyzed for these diverse contexts, we find more similarities in hardware and software design, communications protocols used, connectivity requirements or resilience-building approaches than one might have expected.
Yet research and development and engineering communities working in different CPS contexts are more aware of differences than similarities. Studies conducted in various contexts found limited mutual flow of ideas and best practices among different CPS environments. Numerous factors are responsible for this situation, including operational concerns, traditional work processes in different market segments, confidentiality requirements, differing skills sets in these fields, and many other issues.
While the fragmentation of the field affects the core hardware, software, and communications technologies, it applies especially strongly to additional concerns that need to be considered at the design stage, such as cybersecurity and privacy. Connectivity via communications networks is a recent requirement in many CPS contexts, and many subfields lack expertise in technologies and practices connected to cyber as well as knowledge of technology approaches to fulfill requirements associated with trustworthiness.

The CPS framework has already created a unifying view on the shared model associated with CPS, and, as part of the model, with the CPS trustworthiness. The application of ontologies and reasoning to the space covered by the CPS Framework potentially supports an in-depth analysis that can be formalized for specific contexts, yet is broadly applicable. 

The experimental use case presented in this paper is limited. However, future work based on the same premises will be more extensive. We plan to implement and evaluate parametrization of use cases, test probabilistic models enabled by the same ontologies, and demonstrate more sophisticated reasoning applied to more complex use cases.

\section{Conclusion} 
\label{section:conclusion}

In this paper, we presented a methodology for developing a Conceptual Ontology of the CPS Framework and its Aspects. We then tested parts of such a Conceptual Ontology to illustrate the approach with a use case for CPS, 
the lane keeping/assist scenario of an advanced car. We demonstrated that the model supports multiple aspects of decision making based on the formulation and automatic answering of semantic queries. Although we focused this work on Trustworthiness, the model contains sufficient complexity to demonstrate the capabilities of the approach and its scalability to the full CPS Framework. Our experiment already includes complex considerations such as Transduction and Influence. Our work demonstrates that an ontology-based methodology can aid engineers in identifying and resolving important issues for design, implementation, and validation of CPS.

\noindent \textbf{Acknowledgements.} \\
M. Balduccini was partly supported by NIST grant 70NANB17H260. M. Huth acknowledges the UK EPSRC funded projects EP/N020030/1 and EP/N023242/1.

\end{document}